\documentclass[reprint,amsmath,amssymb,aps]{revtex4-2}
\usepackage{graphicx}
\usepackage{dcolumn}
\usepackage{bm}
\usepackage{xcolor}
\usepackage{comment}
\usepackage{physics}

\usepackage{fancyhdr}
\begin{document}
\fancyhead[R]{\ifnum\value{page}<2\relax\else\thepage\fi}
\title{Grover Search Inspired Alternating Operator Ansatz of Quantum Approximate Optimization
Algorithm for Search Problems}

\author{Chen-Fu Chiang}
 \affiliation{Department of Computer Science, State University of New York Polytechnic Institute, Utica, NY 13502 USA}
 \email{corresponding author: chiangc@sunypoly.edu}

\author{Paul M. Alsing}
\affiliation{Information Directorate, Air Force Research Laboratory, Rome, NY 13441, USA}
\email{paul.alsing@us.af.mil}

\date{\today}

\begin{abstract}
We use the mapping between two computation frameworks
, Adiabatic Grover Search (AGS) and Adiabatic Quantum Computing (AQC), to translate the
Grover search algorithm into the AQC regime. We then apply Trotterization on the schedule-dependent
Hamiltonian of AGS to obtain the values of variational parameters
in the Quantum Approximate Optimization Algorithm (QAOA) framework. The goal is to carry the optimal behavior of Grover search algorithm into the QAOA framework without the iterative machine learning processes.
\end{abstract}

\maketitle
\thispagestyle{fancy}  

\section{Introduction}\label{sec:introduction}
Quantum technologies have advanced dramatically in recent years, both in
theory and experiment. Building a programmable quantum computer involves
multiple layers: algorithms, programming languages, quantum compilers,
efficient decomposition of unitary operators into elementary gates,
the control interface and the physical quantum qubits.
The aforementioned requires much research effort for optimization,
across and within the layers. From a quantum algorithm perspective, even
optimistically assuming the middle layers are perfect, it remains extremely
challenging to use quantum algorithms to solve real-life-size hard problems due
to size limit and errors arising from issues such as precision, random noise and
de-coherence in the quantum devices. Hybrid, short shallow circuit algorithms,
such as Variational Quantum
Eigensolver (VQE) \cite{peruzzo2014variational} and Quantum Approximate
Optimization Algorithm (QAOA) \cite{farhi2014quantum}, are two near-term answers.
From theoretical and implementation aspects, this work aims at
investigating two computation frameworks: Grover search and
QAOA. \\

From a perspective of universal computational models, Quantum Walks (QWs) have become a
prominent model of quantum computation due to their direct relationship to the
physics of the quantum system \cite{farhi1998quantum,kempe2003quantum}.
It has been shown that the QW computational framework is universal for quantum computation
\cite{childs2009universal, lovett2010universal}, and many algorithms now are
presented directly in the quantum walk formulation rather than through a circuit
model or other abstracted method \cite{farhi1998quantum, qiang2016efficient}.
There are multiple quantum computation models,
including the quantum circuit model \cite{shor1998quantum, yao1993quantum,
jordan2012quantum}, topological quantum computation \cite{nayak2008non},
adiabatic quantum computation \cite{mizel2007simple}, resonant transition based
quantum computation \cite{chiang2017resonant} and measurement based quantum
computation \cite{morimae2012blind, gross2007novel, briegel2009measurement,
raussendorf2003measurement}. Notable successes of quantum computation include
Shor’s factoring algorithm \cite{shor1994algorithms} and Grover’s search
algorithm  \cite{grover1996fast}, which manifest indisputable improvements over
the best possible classical algorithms designed for the same purpose. QWs can be
formulated in discrete-time (DTQW) \cite{aharonov1993quantum} and
continuous-time (CTQW) \cite{farhi1998quantum} versions.
It is known that Grover search is a special type of DTQW.
Since both (QAOA and QW) are universal
computational frameworks \cite{childs2009universal, lovett2010universal,
lloyd2018quantum}, there should exist some relationship between those models. One
can extend the techniques from one framework to another by exploring the
connections between the computational models. As the connections are
established, one can further investigate if performance-boosting techniques, such
as spectral gap amplification \cite{somma2013spectral}and catalyst Hamiltonians
\cite{10.5555/2011395.2011396}, can be applied from one framework to the other
to provide additional algorithmic improvement. \\

The structure of this is as follows. The background on AGS, AQC, QAOA is provided
in section \ref{sect:background}. The mapping of Grover to AQC is summarized in
section \ref{sect:ags} and the Trotterization of the schedule dependent AGS to QAOA is
given in section \ref{sect:qaoa_grover}. We give our the error analysis in section \ref{sect:error}.
Finally our conclusion is given in section \ref{sect:conclude}.

\section{Background}\label{sect:background}

\subsection{Adiabatic Quantum Computing}
In the AQC model, $H_0$ is the initial Hamiltonian, $H_f$ is the final
Hamiltonian where the evolution path for the time-dependent Hamiltonian is
\begin{equation}\label{eqn:aqc}
H(s) = (1-s)H_0 + sH_f{\color{black}{,}}
\end{equation}
where $0 \leq s  \leq 1$ is a schedule function of time $t$. The schedule is $s =s(t)$ and
$t$ goes from $0$ to the total run-time $T_a$. The variable $s$ increases at a slow rate
such that the initial ground state evolves and the system state remains as the ground state throughout the adiabatic process. More specifically, the
Hamiltonian at time $t$
\begin{align}
H(s(t))\ket{\lambda_{k,t}} = \lambda_{k,t}\ket{\lambda_{k,t}}{\color{black}{,}}
\end{align}
where $\lambda_{k,t}$ is the corresponding eigenvalue for the
$k${\color{black}{th}} eigenstate $\ket{\lambda_{k,t}}$ at time $t$. For instance,
$\ket{\lambda_{0,t}}$ is the ground state at time $t$. The minimal eigenvalue
gap is defined as
\begin{align}
g = \min_{0 \leq t \leq Ta}(\lambda_{1, t} - \lambda_{0,t}){\color{black}{,}}
\end{align}
where $T_a$ is the total evolution time of the AQC. It is known that $T_a \propto \frac{1}{g^2}$. Let $\ket{\psi (T_a)}$ be
the state of the system at time $T_a$ evolving under the Hamiltonian $H(s(t))$
from the ground state $\ket{\lambda_{0,0}}$ at time $t=0$. The adiabatic theorem
\cite{farhi2000quantum, albash2018adiabatic} states that the final state
$\ket{\psi (T_a)}$ is $\epsilon_1$-close to the real ground state
$\ket{\lambda_{0,T_a}}$ as
\begin{align}\label{eqn:aqc_limit_approx}
|\braket{\lambda_{0, T_a}}{\psi(T_a)}|^2 \leq 1 - \epsilon_1^2,
\end{align}
provided that
\begin{align}
\frac{|\bra{\lambda_{1,t}}\frac{dH}{dt}\ket{\lambda_{0,t}}|}{g^2} \leq \epsilon_1.
\end{align}
\subsection{Quantum Approximate Optimization Algorithm}
QAOA is a promising approach for programming a near-term gate-based hybrid
quantum computer to find good approximate solutions of hard combinatorial
problems. In the near future, the number of reliable quantum gates will be
limited due to noise, de-coherence and scalability.  Due to this fact, hybrid
quantum-classical algorithms have been proposed to make the best of available
quantum resources and integrate them with classical routines. Technically, QAOA
\cite{farhi2014quantum} is a variational ansatz that uses p sets of alternating
non-commuting (Z-basis associated with parameter $\gamma$ and then X-basis
associated with parameter $\beta$) operations on an initial $\ket{+}^{\otimes
n}$ state. With each of the $p$ steps, the state evolves with two unitaries,
\begin{align}
U_t=exp(-i \gamma_t H_c) \textrm{ and }  V_t= exp(-i\beta_tH_0){\color{black}{,}}
\end{align}
where $H_c$ is the cost Hamiltonian of the given optimization (or search)
problem in the computational basis. $H_0 = \sum_i \sigma_i^x$  with $
\sigma_i^x$ is the Pauli X matrix for the $i${\color{black}{th}} qubit. In the
Noisy Intermediate Scale Quantum (NISQ) computing era
\cite{preskill2018quantum}, it is desirable to use shallow circuits to obtain
solutions with high accuracy. Hence, $p$ invocations of the QAOA operator would be
\begin{align}\label{eqn:qaoa_2op}
U_{qaoa} = \prod_{t=1}^p V_t U_t{\color{black}{,}}
\end{align}
and $p$ is expected to be some small number to avoid unnecessary decoherence.
QAOA aims at solving optimization problems with a short circuit and provides
acceptable approximate solutions. Numerous studies have been conducted to find
optimal $\beta, \gamma$ for each of the step for a shorter circuit
and benchmark the performance of QAOA
\cite{wecker2016training,crooks2018performance, zhou2018quantum,
willsch2019benchmarking}. In practice, in order to find the optimal values of
$\beta_i$ and $\gamma_i$ parameters, several iterations of
optimization are required as illustrated in Fig. 1
\cite{PhysRevX.10.021067}.
The final state is measured to obtain expectation value with respect to the
objective function $H_c$. Then the result is fed to the classical optimizer.
Based on the condition one sets, normally the iteration stops when no more
improvement can be found or the improvement is negligible. However, the post-processing of optimization to find proper values of $\beta_{i+1}$ and $\gamma_{i+1}$ from the $i${\color{black}{th}} iteration using machine learning might be costly.

\begin{figure}[htbp]
\centerline{\includegraphics[scale=.25]{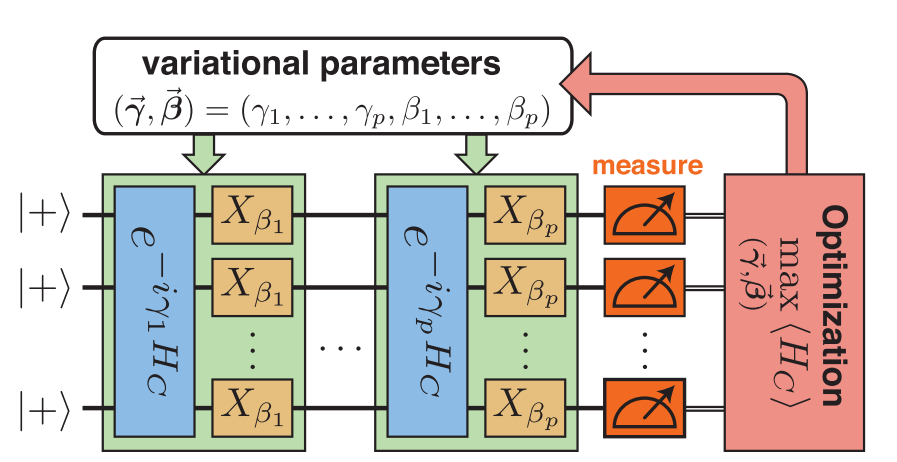}}
{\cite{PhysRevX.10.021067} {\color{black}{Fig.}} 1: QAOA with classical optimizer to find the optimal parameters of $\{\gamma_i, \beta_i\}$ as $\{\vec{\gamma}, \vec{\beta}\}$.}
\label{fig:qaoa}
\end{figure}

In the following sections we explore the connection between QAOA and AGS. Here AGS refers to applying the AQC time-dependent Hamiltonian approach to Grover's problem searching for a marked element in the unstructured database. An unstructured search problem of size $N = 2^n$ where the adjacency matrix $A$ has ones everywhere except all zeros on the diagonal. Let $\ket{\omega}$ be the target state while $\ket{s} = \ket{+}^{\otimes n}$ be the initial uniform superposition state. Our initial investigations on the process of mapping CTQW and AGS to QAOA via AQC can be can be viewed as Fig. 2.
{\color{black} {The Grover inspired approach AGS was first introduced in \cite{roland2002quantum}. There are two paths as shown in Fig. 2, one is CTQW-AQC-QAOA while the other is AGS-AQC-QAOA. This work is mainly for the AGS-AQC-QAOA path. The AGS based path does not cause irreconcilabilities but the CTQW based path did \cite{wong2016irreconcilable}.} A more detailed explanation regarding the irreconcilability and potential solutions is at the appendix \ref{sect:appdixA} for the CTQW-AQC-QAOA path. }

\begin{figure*}[htbp]
\centerline{\includegraphics[scale=.8]{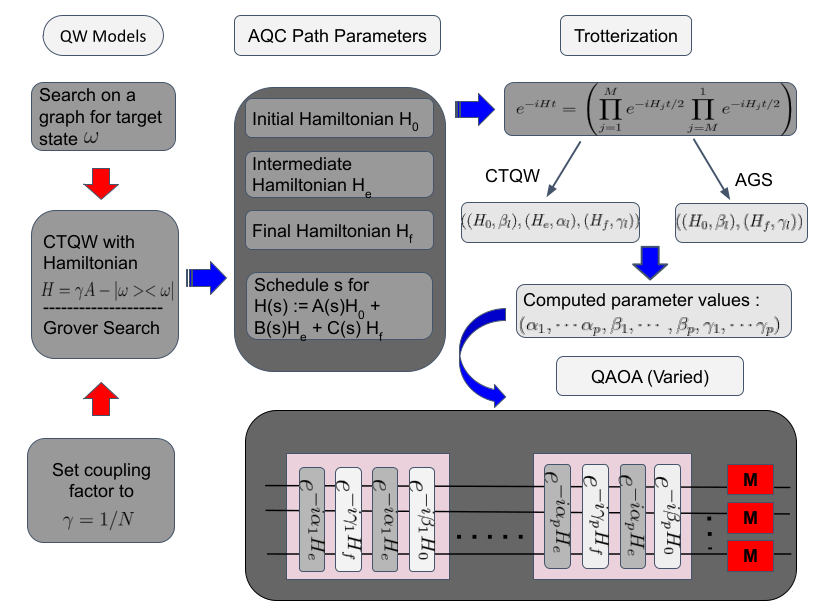}}
{{\color{black}{Fig.}} 2:The process of translation from QW based framework to QAOA to obtain the values of variational parameters in QAOA. {\color{black}One path is CTQW-AQC-QAOA while the other is AGS-AQC-QAOA.}}
\label{fig:mappingtoqaoa}
\end{figure*}

\section{Mapping: Grover Search to AQC}\label{sect:ags}
{\color{black} The time-dependent Hamiltonian approach \cite{roland2002quantum} was applied to the Grover’s search, searching a marked item in an unstructured database. Grover’s algorithm was originally presented as a discrete sequence of unitary logic gates. In \cite{roland2002quantum} it turns to another type of quantum computation where the state of the quantum register evolves continuously under the influence of some driving Hamiltonian. By adjusting the
evolution rate of the Hamiltonian so as to keep the evolution adiabatic on each infinitesimal time
interval, the total running time is of order $\sqrt{N}$ where $N$ is the number of items in the database.}
Let the time-dependent Hamiltonian as defined in Eqn. (\ref{eqn:aqc}) with the
initial Hamiltonian $H_0$ and the final Hamiltonian $H_f$ being
\begin{align}
H_0 = I - \dyad{\psi_0}{\psi_0}, \quad  H_f =I - \dyad{\omega}{\omega}.
\end{align}
By such a setting, the initial uniform superposition state $\ket{\psi_0}$ is the
ground state for $H_0$ and the target state $\ket{\omega}$ is the ground state
for $H_f$. The system state evolves in the $\{\ket{\omega}, \ket{r}\}$ basis
with the Hamiltonian as \cite{wong2016irreconcilable}
\begin{align}\label{eqn:grover_hs}
H(s) =
\begin{pmatrix}
(1-s)\frac{N-1}{N} & -(1-s)\frac{\sqrt{N-1}}{N} \\
-(1-s)\frac{\sqrt{N-1}}{N} & 1 - (1-s)\frac{N-1}{N} \\
\end{pmatrix}.
\end{align}

\noindent
{\color{black} Following the adiabatic theorem, at any time $t$ during the evolution,
\begin{equation}\label{eqn:epsilon}
|\braket{E_0(t)}{\psi_0(t)}|^2 \ge 1-\epsilon_1^2{\color{black}{,}}
\end{equation}
where $E_0(t)$ is the true ground state and $\psi_0(t)$ is the system state following the adiabatic evolution. Instead of using a linear evolution of $s(t)$, in \cite{roland2002quantum} it adapts the evolution $ds/dt$ to the local {\color{black}{adiabaticity}} condition.
It is therefore
\begin{equation}\label{eqn:dsdt}
|\frac{ds}{dt}| = \epsilon_1 g^2(t){\color{black}{,}}
\end{equation}
where $g(t)$ is the energy gap of the system at time $t$. The running time $t$ is a function of the interpolation schedule $s$ such that}
\begin{align}
t = & \frac{N}{2\epsilon_1 \sqrt{N-1}}\Big\{\arctan(\sqrt{N-1}(2s-1)) \nonumber \\
      & + \arctan(\sqrt{N-1})\Big\}{\color{black}{,}}
\end{align}
and the schedule satisfies the adiabatic theorem. {\color {black}It is further shown that the running time is optimal as $T = O(\sqrt{N})$ when $s=1$ \cite{roland2002quantum}.} By the trigonometry formula {\color{black}{$\arctan(x) + \arctan(y) = \arctan({\frac{x+y}{1-xy}})$} mod $\pi$} and
the approximation $N \simeq N-1$ when $N$ is large, we know that the angle
for the tangent function is bounded as
\begin{align}
0 \leq \frac{2t\epsilon_1 \sqrt{N-1}}{N} \leq \pi,
\end{align}
which echoes the fact that $0 \leq t \leq T$ when $\epsilon_1$ is some negligible constant. The interpolate schedule $s$
with respect to time $t$ is
\begin{align}\label{eqn:ags_s}
s 
   &=  \frac{\sqrt{N} \tan(\frac{2t\epsilon_1}{\sqrt{N}})}{2(\sqrt{N} \tan(\frac{2t\epsilon_1}{\sqrt{N}}) +1)}.
\end{align}
For simplicity, let us call this adiabatic local search based on Grover as
Adiabatic Grover Search (AGS).

\section{Connection: AGS to QAOA via AQC}\label{sect:qaoa_grover}
For simplicity, let us define the time-dependent Hamiltonian $H(s)$ in
{\color{black}{Eq.}} (\ref{eqn:grover_hs}) as
\begin{align}\label{eqn:grover_aqc_sim}
H(s) = A(s)H_0 + C(s) H_f.
\end{align}
To reflect {\color{black}{Eq.}} (\ref{eqn:grover_aqc_sim}) in AGS, one can decompose
the evolution operator into some large $R$ steps using Suzuki-Trotter formula,  the state evolution
of the system is
\begin{align}\label{eqn:grover_aqc_state}
U(T) &= exp[-i \int_{{\color{black}{0}}}^{{\color{black}{T}}} H(t) dt].
\end{align}
The Suzuki-Trotter decomposition states $e^{A+B} = \lim_{R \rightarrow \infty} (e^{A/R}e^{B/R})^R$ and let us choose $\tau = T/R$. Since the system Hamiltonian evolves based on the schedule $s$, we can further write
\begin{align}\label{eqn:grover_aqc_decomp}
U(T)
& \simeq \prod_{l=1}^R exp [-iH(s_l)\tau]\\
& = \prod_{l=1}^R exp[-i\big(A(s_l)H_0 + C(s_l)H_f\big)\tau] \nonumber \\
& = \prod_{l=1}^R \big(e^{-i\tau A(s_l)H_0/2}e^{-i\tau C(s_l)H_f}e^{-i\tau A(s_l)H_0/2}\big){\color{black}{,}} \nonumber
\end{align}
by using the second order Trotter method where $s_l$ is obtained using {\color{black}{Eq.}} (\ref{eqn:ags_s})
\begin{align}\label{eqn:sl_qg}
s_l = \frac{\sqrt{N} \tan(\frac{\pi l}{R})}{2(\sqrt{N} \tan(\frac{\pi l }{R})+1)}{\color{black}{,}}
\end{align}
and $t = \frac{lT}{R}$ and $T \simeq O(\frac{\pi\sqrt{N}}{2\epsilon_1})$.
To map to QAOA $U_{qaoa}$ operator with $p$ steps where
\begin{align}\label{eqn:qaoa_grover}
U_{qaoa} = \prod_{l=1}^p V_l U_l = \prod_{l=1}^p (e^{-i \beta_l H_0}e^{-i\gamma_l H_f}),
\end{align}
one can neglect $e^{-i\tau A(s_1)H_o/2}$ because its action on
$\ket{+}^{\otimes N}$ yields only a global phase factor. By matching
{\color{black}{Eq.}} (\ref{eqn:grover_aqc_decomp}), one sets $p=R$ and obtains
\begin{align}\label{eqn:coefficients}
\gamma_{l\in \{1, 2, \cdots, R\}}  &= \tau C(s_l){\color{black}{,}} \\
\beta_{l \in \{1, 2, \cdots, R-1\}} &= \tau (A(s_l) + A(s_{l+1}))/2 {\color{black}{,}}\\
\beta_R  & =  \tau A(s_R)/2 {\color{black}{.}}
\end{align}
For AGS, the schedule follows Eqn. (\ref{eqn:aqc}) as
\begin{align}
A(s) = (1-s), \quad
C(s) = s {\color{black}{,}}
\end{align}
and we will obtain
\begin{align}\label{eqn:coef_qg}
\gamma_{l\in \{1, 2, \cdots, R\}} &=  \tau s_l {\color{black}{,}}\\
\beta_{l \in \{1, 2, \cdots, R-1\}} &= (\tau/2)\big(2- (s_l + s_{l+1})\big) {\color{black}{,}}\\
\beta_R &= (\tau /2)(1-s_R).
\end{align}

\section{Errors}\label{sect:error}
The approximation error from the translation between models is two-fold: one comes from the AQC simulation error
$\epsilon_1$ as indicated in {\color{black}{Eq.}} (\ref{eqn:aqc_limit_approx}) and the other
source of error, $\epsilon_{2k}$, is from the Hamiltonian simulation via
Trotterization. Let the
approximated unitary be $\tilde{U}$ and the expected total error be bounded from
above by $\epsilon$, we have
\begin{align}\label{eqn:total_error}
\left\Vert U - \tilde{U} \right \Vert = \epsilon_{2k} + \epsilon_1  = \epsilon.
\end{align}
Now we need to investigate the value of Trotterization
steps $R$ to obey the desired $\epsilon$ error in the simulation. For even
higher-order, let us denote it as $2k$-th order for $k>0$,  the formula can be
constructed recursively and $U_{2k}(t)$  is of
the form \cite{suzuki1992general}
\begin{align}
& [U_{2k-2}(s_kt)]^2U_{2k-2}((1-4s_k)t)[U_{2k-2}(s_kt)]^2 \nonumber \\
&= e^{-iHt} + O((Mt)^{2k+1}/R^{2k}),
\end{align}
where $H =\sum_{j=1}^M H_j$ and $s_k = 1/(4-4^{1/(2k-1)})$. In general, via the
above form, arbitrary high-order formulas can be constructed. But in practice the
fourth order ($2k=4$) is preferred for most practical problems as the cost from
constructing more complex higher-order operators would offset the benefits of
Trotterization. To confine the simulation error stated in {\color{black}{Eq.}}
(\ref{eqn:total_error}),  we must satisfy the condition that $\epsilon_{2k} \leq
\epsilon - \epsilon_1$. This immediately shows that $R$ should be chosen
accordingly, as listed in Table \ref{tbl:1}.

\begin{table}[hbt]
\centering
\begin{tabular}{|c|c| }
\hline
{\bf Variable}       & {\bf AGS} \\ \hline \hline
$\epsilon_{2k}$
& $ O((2t)^{2k+1}/R^{2k})$    \\ [2ex]\hline
R
&  $O\left( \sqrt[2k]{\frac{(2t)^{(2k+1)}}{(\epsilon - \epsilon_1)}}\right)$ \\ \hline
\end{tabular}
\caption{The Hamiltonian simulation error $\epsilon_{2k}$ and corresponding discrete-time steps $R$.}
\label{tbl:1}
\end{table}

Recalling that adiabatic Grover have the optimal run-time
$O(\sqrt{N})$, we have to  set  $t =  O(\sqrt{N})$. When using Trotterization
for Hamiltonian simulation, if $\epsilon_{2k}$ is some small  constant, we can
conclude that at $k=1$, the required discrete-time step is sub-optimal as $R
\simeq O(N^{3/4})$.  As $k$ increases, $R$ approaches $O(\sqrt{N})$. In our case
for variational variable values based on the second order ($k=1$) approximation,
the corresponding QAOA should obtain an $\epsilon$-close solution with the
sub-optimal running time $R=O(N^{3/4})$.

\section{Conclusion and Future Work}\label{sect:conclude}
In this work, we explore ways to let QAOA simulate the behavior of optimal
search by AGS. The motivation is to find the values of variational
parameters from a theoretical approach, instead of heuristic
approaches. We discover the values of the variational parameters by letting QAOA simulate AGS via AQC. The AGS obeys the conventional AQC
and the mapping is straightforward. Finally, from an error control
perspective, to achieve the same degree of accuracy $\epsilon$, both mappings indicate they have the same number of Trotterization steps in the big O notation. \\

For future investigation, we consider the connection CTQW to QAOA to be another interesting direction. There are several variations of AQC to improve the performance. The variations
are based on modifying the initial Hamiltonian and the final Hamiltonian
\cite{10.5555/2011395.2011396, perdomo2011study} or adding a catalyst
Hamiltonian $H_e$ \cite{10.5555/2011395.2011396}. The catalyst vanishes at the initial and the final times but is present at intermediate times. For instance, a conventional catalyst assisted AQC is expressed as
\begin{align}\label{eqn:typical_catalyst_aqc}
H(s) = (1-s)H_0 + s(1-s)H_e + sH_f.
\end{align}
The form of $H_e$ is important but even a randomly chosen catalyst can help
in improving run time \cite{10.5555/2011395.2011396, zeng2016schedule}. The use of catalyst Hamiltonian also suggests an additional variational parameter $\alpha$ is needed when Totterizing to QAOA as shown in Fig. 2.
\begin{acknowledgments}
CC gratefully acknowledges the support from the seed grant funding
from the State University of New York Polytechnic Institute and the support from the Air Force Research Laboratory Summer Faculty Fellowship Program (AFSFFP). PMA would like to acknowledge support of this work from
the Air Force Office of Scientific Research (AFOSR). Any opinions, findings and conclusions or recommendations
expressed in this material are those of the author(s) and do not
necessarily reflect the views of Air Force Research Laboratory.
\end{acknowledgments}
{\color {black}{
\appendix
\section{CTQW-AQC-QAOA Path}\label{sect:appdixA}
\subsection{Concern}
To construct the time-dependent Hamiltonian $H(t)$, one can use the results from
\cite{wong2016irreconcilable} where the AQC search follows the CTQW search on a
complete graph. By
choosing the coupling factor $\gamma = 1/N$ and letting $\ket{r}$ be the uniform
superposition of non-solution states such that
\begin{align}
\ket{r} = \frac{1}{\sqrt{N-1}}\sum_{i\neq \omega} \ket{i},
\end{align}
the resulting Hamiltonian
\cite{farhi1998analog} for CTQW in
the $\{\ket{\omega}, \ket{r} \}$ basis
is
\begin{equation} H = \frac{-1}{N}
\begin{pmatrix}
N+1 & \sqrt{N-1} \\
\sqrt{N-1}& N -1 \\
\end{pmatrix}.
\end{equation}

\noindent
The system state $\ket{\psi(t)}$ evolution path following the unitary $e^{-iHt}$ is considered
as the ground state for the adiabatic path. The time-dependent
Hamiltonian $H(t)$ with $\ket{\psi(t)}$ as ground state is \cite{wong2016irreconcilable}
\begin{align}\label{eqn:ctqw_aqc}
H(s) =& \sqrt[4]{\frac{s(1-s)}{4\epsilon_1^2N}}[(1-s)H_0 + \sqrt{s(1-s)}H_e  \nonumber \\
           &+ s H_f],
\end{align}
where the schedule $s(t) = sin^2(\frac{t}{\sqrt{N}})$. The initial,
catalyst, and final Hamiltonians in the $\{\ket{\omega}, \ket{r} \}$ basis
are
\begin{align} \label{eqn:Hamiltonians}
H_0  &=
\begin{pmatrix}
\frac{N-2}{N} & -2\frac{\sqrt{N-1}}{N} \\
-2\frac{\sqrt{N-1}}{N} & -\frac{N-2}{N} \\
\end{pmatrix},  \\
H_e & =
\begin{pmatrix}
0 & -2i\sqrt{\frac{N-1}{N}} \\
2i\sqrt{\frac{N-1}{N}}  & 0\\
\end{pmatrix},
H_f  =
\begin{pmatrix}
-1 & 0 \\
0  & 1\\
\end{pmatrix}. \nonumber
\end{align}
The Hamiltonians can be further written as \cite{wong2016irreconcilable}
\begin{align} \label{eqn:Hamiltonians_simple}
H_0 &= \dyad{\psi_0^{\perp}}{\psi_0^{\perp}} - \dyad{\psi_0}{\psi_0}, \quad
H_f = \dyad{r}{r} - \dyad{\omega}{\omega}, \nonumber \\
H_e &=2i\sqrt{\frac{N-1}{N}}( \dyad{r}{\omega} - \dyad{\omega}{r}),
\end{align}
where $\ket{\psi_0}$ is the initial uniform superposition state.
In comparison to the typical expression in {\color{black}{Eq.}} (\ref{eqn:aqc}), the
intermediate extra Hamiltonian $H_e$ in {\color{black}{Eq.}} (\ref{eqn:ctqw_aqc})
facilitates the driving between $\ket{\omega}$ and $\ket{r}$. However, there
exists irreconcilability in the CTQW-inspired AQC path.

In Eqn.(\ref{eqn:ctqw_aqc}), the following parameters were computed during the mapping \cite{wong2016irreconcilable}:
\begin{itemize}
    \item the scaling factor $\propto \sqrt[4]{\frac{1}{N}}$ of $H(s)$,
    \item the corresponding catalyst Hamiltonian $H_e$ provides power greater than a typical Yes/No oracle,
    \item the coefficient function of catalyst Hamiltonian $H_e$ as $\sqrt{s(1-s)}$.
\end{itemize}

The main concerns arising from {\color{black}{Eq.}} (\ref{eqn:ctqw_aqc}) are twofold. One concern is the factor $ \sqrt[4]{\frac{s(1-s)}{4\epsilon_1^2N}}$ of $H(s)$.
The {\color{black}{adiabatic theorem}} \cite{griffiths2018introduction} states that if we prepare system at time $t=0$ in its ground state and let it evolve under the
Hamiltonian $H(t)$, the system achieve a fidelity of $1-\epsilon_1$ to the target state, provided that
\begin{equation}\label{eqn:dhdtgmin}
\frac{|\langle\frac{dH}{dt}\rangle_{0,1}|}{g_{min}^2} \leq \epsilon_1 \text{,  where } g_{min} = \min_{0\leq t \leq T}E_1(t) - E_0(t) .
\end{equation}
Here $\langle\frac{dH}{dt}\rangle_{0,1}$ are the matrix elements of $dH/dt$ between the two corresponding eigenstates.
$E_0(t)$ and $E_1(t)$ are the ground energy and the first excited energy of the system at time $t$. Given the $H(s)$ in
 {\color{black}{Eq.}} (\ref{eqn:ctqw_aqc}), one might conclude that a factor of  $ O(\sqrt[4]{1/N})$ significantly reduces the required time to achieve $1-\epsilon_1$ precision.
This might be misleading as the $g_{min}$ of $H(s)$ also carries that factor.

\noindent
The second concern is that the catalyst $H_e$ provides power
greater than a typical Yes$\slash$No oracle as it maps non-solution states
to a solution state and a solution state to non-solution states.
Provided initially the we start with a superposition state with amplitude of
 $\sqrt{\frac{N-1}{N}}$, it takes time of $O(1)$ for this catalyst
to drive to the solution state from the initial state.

\subsection{Strategy} \label{sect:m-qw-aqc}
To address the irreconcilability issue, we propose to (1) drop  factor $ \sqrt[4]{\frac{s(1-s)}{4\epsilon_1^2N}}$ and (2) modify the catalyst Hamiltonian $H_e$ to be a regular oracle.
The form of $H_e$ is important but even a randomly chosen catalyst can help in improving run time \cite{10.5555/2011395.2011396, zeng2016schedule}. Aiming for being optimal in the translated algorithm without considering other constraints,
CTQW-inspired {\color{black}{adiabatic}} path has $H_e = 2\sqrt{\frac{N-1}{N}} i XZ$ in the $\{\ket{\omega},  \ket{r}\}$ basis that
provides more power than a standard {\color{black}{oracle}}. To avoid disputes, we will drop the imaginary number $i$ and the $X$ operator. The
$Z$ alone behaves as a conventional Yes \slash No {\color{black}{oracle}}. A slight difference is that since we are in the $\{\ket{\omega},  \ket{r}\}$ basis (not $\{ \ket{r},\ket{\omega}\}$ basis),
the $Z$ operator behaves like the conventional {\color{black}{oracle}} with an additional minus sign. Let $M$ be the magnitude scalar equal to $2\sqrt{\frac{N-1}{N}}$ computed from CTQW.
The new modified $H_m(s)$ schedule is therefore defined as
\begin{align}\label{eqn:ctqw_m_aqc}
H_m(s) =&(1-s)H_0 + f_{z}(s)MZ + s H_f,
\end{align}
where $f_{z}(s)$ is our chosen $s$-dependent coefficient for catalyst $Z$. In addition to $f_{z}(s) = \sqrt{s(1-s)}$ in \cite{wong2016irreconcilable},
functions that reach its maximum when $s=1/2$ would be good candidates for $f_{z}(s)$. For instance, another good candidate is $f_{z}(s)=\frac{\sin(s \pi)}{2}$.
{\color{black}{Because of $0\leq \sqrt{s(1-s)}\leq 1/2, 0 \leq \sin(s \pi) \leq 1$ as $0\leq s \leq 1$, one has}} to use $\frac{1}{2}$ on the sine function  is to offset the magnitude $M$
to bound the norm of $H_e$. Based on the modified schedule, oracle-like catalyst Hamiltonian, one can run the experiment by simulation for CTQW-AQC part using {\color{black}{Eq.}} (\ref{eqn:ctqw_m_aqc}) and compare the result with the optimal AGS-AQC part using {\color{black}{Eq.}} (\ref{eqn:grover_hs}). Our initial investigation indicates the modified CTQW-AQC, bypassing the irreconcilability, remains optimal as the AGS-AQC did. }}

\bibliography{ref}
\end{document}